\documentclass[11pt,preprint]{aastex}
\begin{document}

\title{INTRIGUING REVELATIONS FROM LITHIUM, BERYLLIUM, AND BORON}

\author{Ann Merchant Boesgaard}
\affil{Institute for Astronomy, University of Hawai`i at M\-anoa, \\ 2680
Woodlawn Drive, Honolulu, HI {\ \ }96822 \\ } 

\email{annmb@hawaii.edu}

\begin{abstract}
This is a report on some highlights of research on the rare light elements,
lithium (Li), beryllium (Be), and boron (B), that I presented in my Henry
Norris Russell Lecture in January, 2020.  It is not a comprehensive review of
work on these light elements, but contains sections on Big Bang
nucleosynthesis of Li and the rarity of these light elements.  It includes
information on how they are observed, both historically and currently, and the
difficulties entailed in determining their abundances.  The production of Li,
Be, and B is ongoing so the youngest stars contain the most Li in their
atmospheres, and they have had less time to destroy it.  All three elements
are readily destroyed in stellar interiors, but have differing degrees of
susceptibility to the particular nuclear fusion reactions which deplete their
surface content.  This feature makes them remarkably good probes into the
otherwise unobservable interiors of stars and provides insights into internal
mixing processes.  It also enhances the use of two or more of the three in
sorting out the various processes at work in the insides of stars.
\end{abstract}

\section{INTRODUCTION}

Jesse Greenstein introduced me to the delights of stellar spectroscopy during
a summer job at Caltech before my graduate school days at the University of
California at Berkeley.  We were studying the Ba II star, $\zeta$ Cap, which
was rich in spectral lines of many rare earth elements.  This interest
continued through a UCB graduate course taught by George Wallerstein where I
did a term paper on the rare light elments.  My first published research
project was an attempt to determine the Li isotope ratio in two Hyades F
dwarfs; there was no clear evidence of $^6$Li (Merchant et al.~1965).  George
Herbig, my mentor and Ph.D thesis advisor, guided me through a research
project on Be in F and G dwarfs (Merchant 1966).  He called my attention to
the strong Li features in M giants and supergiants which evolved into my
thesis research.  Lithium was present in all the 58 stars we studied at high
dispersion but showed a large range of 250 in Li abundance in those cool
evolved stars (Merchant 1967).  This range could be attributed in part to the
dilution of the surface Li as the outer convection zone deepened in evolved
giant stars as had just been proposed by Iben (1965).

In the course of my post-doctoral fellowship at Caltech with Jesse Greenstein,
I moved up the periodic table and studied the isotopes of Mg in 10 cool stars
via the MgH features (Boesgaard 1968).  Work on the Ti/Zr ratio, even further
along on the periodic table, on cool stars was completed then (Boesgaard
1970a).  But Li was not forgotten in work on Li in heavy metal stars and with
it was the discovery of the super Li-rich S star, T Sgr (Boesgaard 1970b).

In 1967 I accepted a position as an Assistant Professor at the University of
Hawaii (UH).  Work on constructing the NASA-funded 88-inch (2.2m) telescope
was underway at that time on Mauna Kea and it was ready for spectroscopic work
in 1969.  The superior quality of Mauna Kea for astronomical observations
became clearly apparent in site surveys before the site selection and in
subsequent astronomical work.  The NASA Planetary Patrol photographs from the
24-inch telescope on Mauna Kea produced the finest images of all the
comparable telescopes placed at observing sites around the world.  By 1979
there were dedications of three more significant telescopes on Mauna Kea:
Canada-France-Hawaii Telescope (3.6 m), NASA Infrared Telescope (3m), and
United Kingdom Telescope (3.8 m).  By the 1990s there were four still larger
telescopes: the two 10-m Keck I and Keck II telescopes, the 8-m Subaru
Japanese National telescope and the AURA 8-m Gemini North telescope.  No
telescopes have been placed on the actual summit of Mauna Kea, the highest
point in the Pacific Ocean.

\begin{figure}[htb!]
\epsscale{0.6}
\plotone{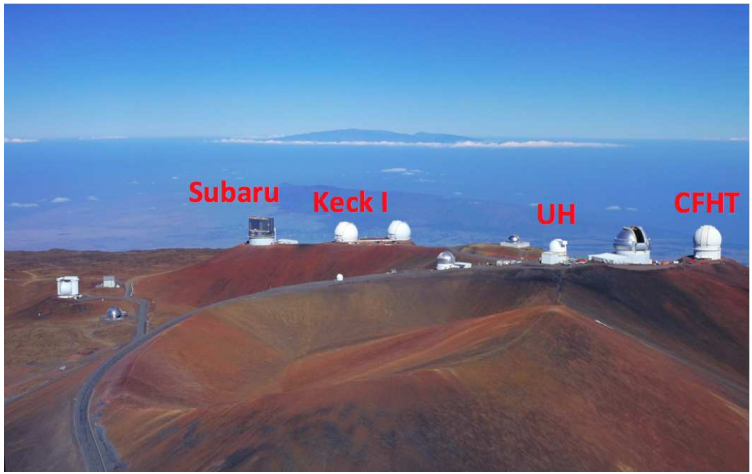} 
\caption{The Mauna Kea observatories showing the domes of the optical, infrared
and sub-millimeter telescopes.  The telescopes labeled are the ones used for
this work.  The astronomy precinct is 525 acres located within the 11,228 acre
science reserve at the top of this massive mountain.  The actual summit of
Mauna Kea can be seen in the foreground on the right - in front of the UH dome
- in this aerial view.  That summit itself is free of any telescopes.  The
island of Maui with its ring of trade wind clouds can be seen in the distance
with the summit of Haleakala.  Original photograph taken by R. J. Wainscoat.}
\end{figure}

Figure 1 shows an aerial photograph of the top of Mauna Kea.  The four
telescopes with which I took most of my spectra for work on  the light elements
are labeled.  Mauna Kea is a massive mountain extending 14,000 ft (4200 m)
above sea level and 17,000 ft below the surface of the Pacific Ocean.  Mauna
Kea is the best observing site on the planet for multiple reasons.  Many are
due to its location far from any land mass in the middle of the ocean: dark
skies, clean air, with almost no light pollution.  Honolulu is below the
horizon.  Its high elevation puts it above the inversion layer resulting in
clear skies, low humidity and low water vapor.  The weather is good year round
and the nighttime hours are long year round.  The astronomical seeing is
exceptional.  In addition the low latitude of the site at +20 N means that
nearly all of the sky can be observe from Hawaii.  The Mauna Kea Science
Reserve is an area of 11,288 acres near the summit and the ``Astronomy
Precinct'' is a small part of that at 525 acres.

\section{BIG BANG NUCLEOSYNTHESIS PRODUCES $^7$Li}

In a hot Big Bang Universe nucleons are a trace component among photons and
 neutrinos.  The epoch of nucleosynthesis begins with the formation of the
hydrogen isotope, $^2$H, then $^4$He, followed by $^3$H and $^3$He at the
expense of the neutrons.  The formation of these nuclei can be seen in Figure
2 starting with the protons and neutrons at the left.  

\begin{figure}[htb!]
\epsscale{1.0}
\plotone{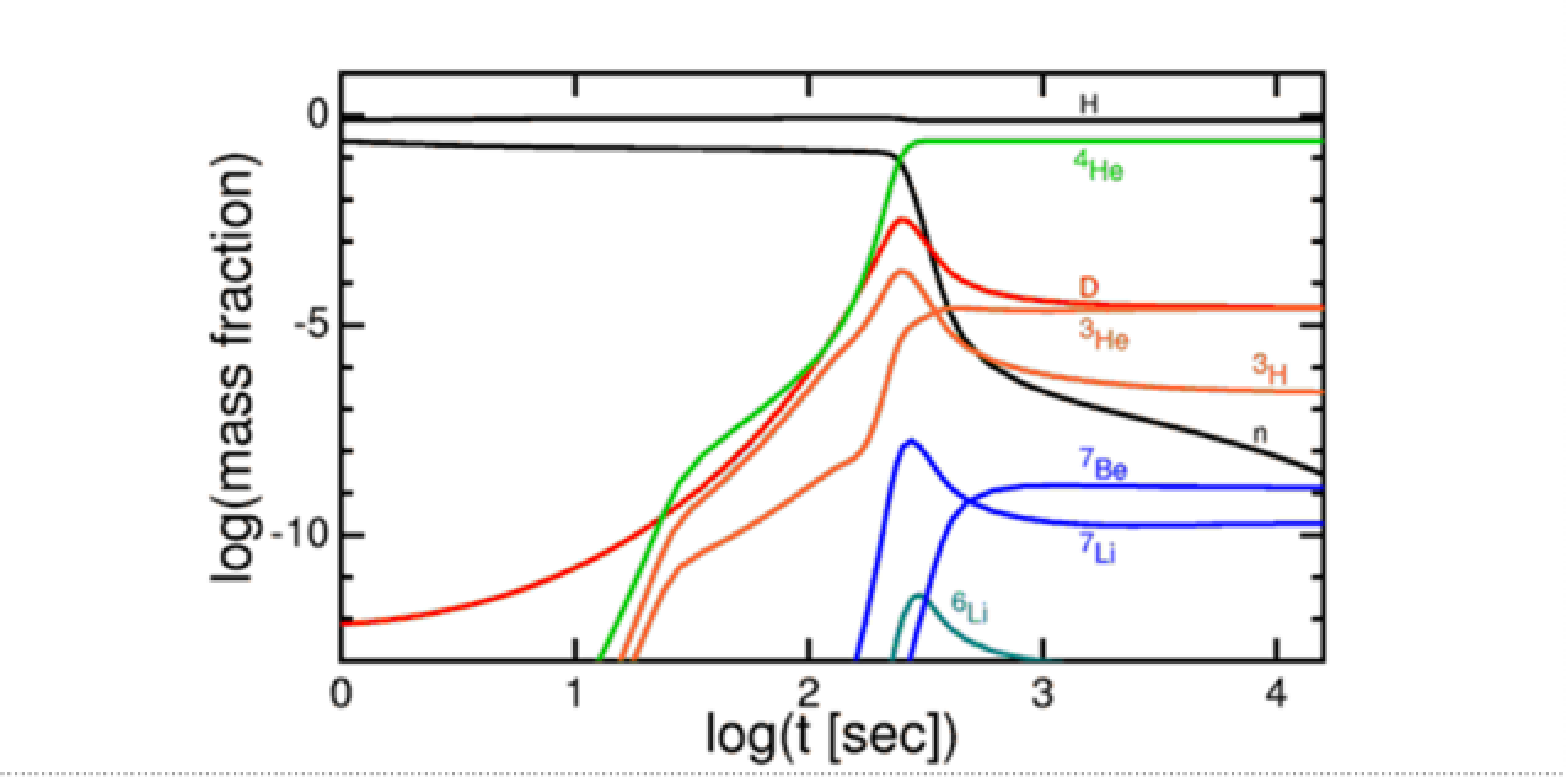}
\caption{Element synthesis in the first 15 minutes after the Big Bang, showing
the mass fraction of each component with time in seconds.  (This
figure is based on data from Burles, Wollett \& Truran (1990) and created by
E.L. Wright at UCLA).}
\end{figure}

In addition some $^7$Li is formed during this element synthesis.  The amount
of $^7$Be that does form decays into $^7$Li.  The reactions are shown
pictorially in Figure 3.  So H, He, and some Li are formed in the Big Bang,
but {\bf nothing else}.  It is important for the rest of the history of the
universe -- and for us -- that there is \underline{no stable mass 5} and
\underline{no stable mass 8}.  Mass 5 would result from the fusion of a H
nucleus with a helium nucleus.  Mass 8 would occur from the fusion of 2 He
nuclei, but $^8$He decays back to 2 He nuclei.  (The exception is inside stars
when the temperature is hot enough to fuse that $^8$He nucleus immediately
with $^4$He to make an excited state of $^{12}$C in the triple-alpha process.)

The periodic table of the elements shows how very different the universe is
today.  Once stars are formed more elements were created.  It is the synthesis
of the elements in stars that gets the universe from the Big Bang to today.
Hydrogen and helium still dominate by orders of magnitude.  They are followed
by C, N, and O, completely skipping over Li, Be, and B.

\begin{figure}[htb!]
\epsscale{0.55}
\plotone{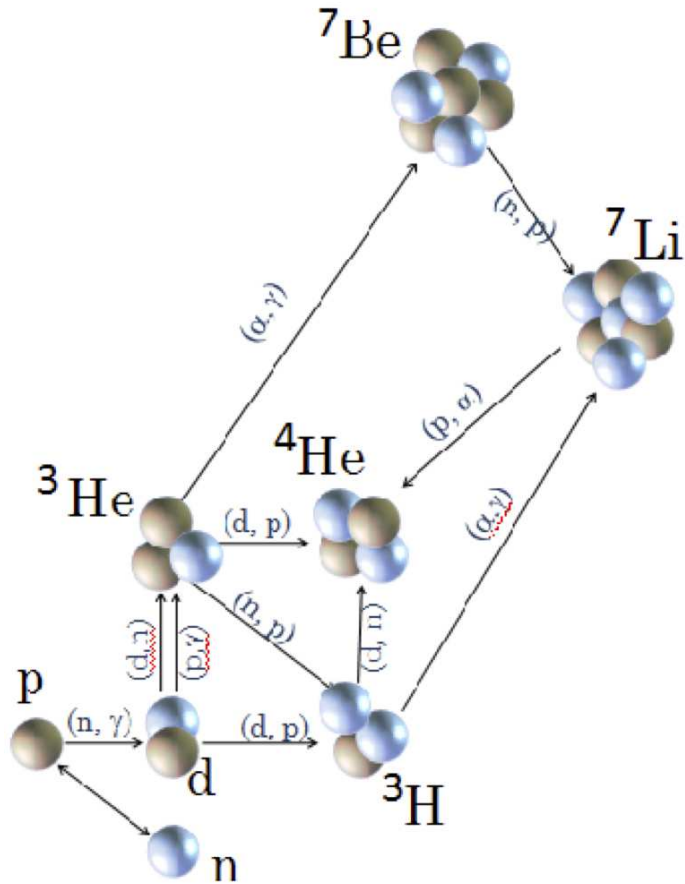} 
\caption{Nuclear reactions that occur with protons and neutrons in the first
15 minutes of the Universe.  It shows how $^7$Li is formed as well as the decay
of $^7$Be to $^7$Li.}
\end{figure}

\section{LITHIUM, BERYLLIUM AND BORON AND THEIR RARITY}

A significant and far-ranging research paper was produced by Burbidge,
Burbidge, Fowler \& Hoyle (1957), commonly known as B$^2$FH, on the
``Synthesis of the Elements in Stars.''  They characterized element formation
by several processes: H-burning, He-burning, $\alpha$-process, e-process,
s-process, r-process, p-process, and the x-process.  There the x-process,
where the x stands for ``unknown,'' was for the light elements, Li, Be, and B.
One example of the distribution of elements in the Sun/solar system by Anders
\& Grevesse (1989) is shown in Figure 4.  This demonstrates clearly how rare
those three light elements are relative to their neighbors on the periodic
table: H and He on the low side and C, N, O all the way past Fe on the heavier
side.

\begin{figure}[htb!]
\epsscale{1.0}
\plotone{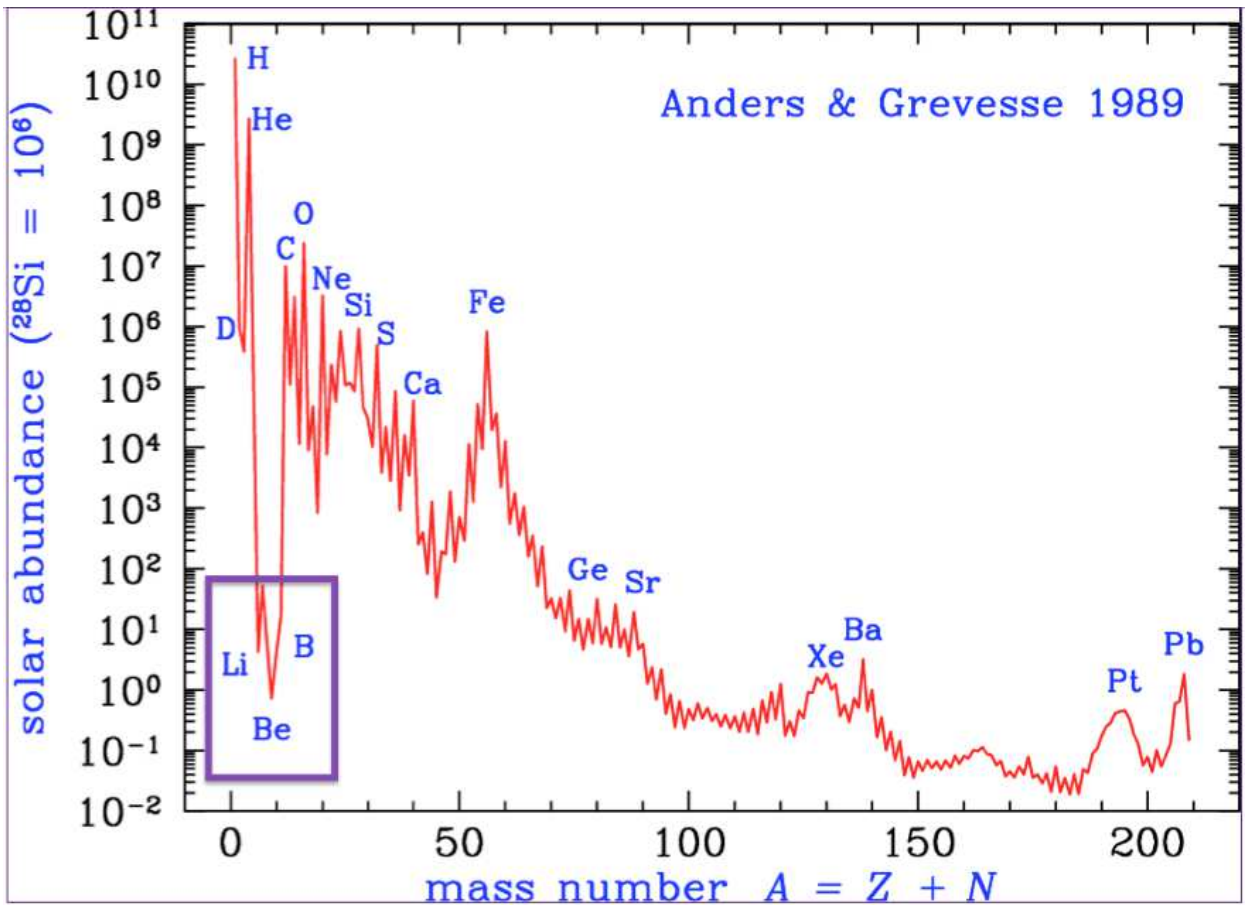}
\caption{Solar/solar system abundance distribution by atomic mass from Anders
\& Grevesse (1989).  The light elements Li, Be, and B, seen at the left, are
very low in abundance compared to their neighbors.  Data from Anders \&
Grevesse (1989).}
\end{figure}

A process to create Li, Be, B could not be sustained by nuclear fusion
reactions inside stars.  The light elements could be formed there but would be
destroyed there and in the cooler, upper layers of the star, $\sim$2.5 x
10$^6$ K for Li, $\sim$3.5 x 10$^6$ K for Be and $\sim$5 x 10$^6$ K for B.
Another environment was required.  An early possibility was in the
interstellar gas where energetic protons and neutrons bombard abundant atoms
of C, N, O and break them into smaller pieces including Li, Be and B.  This
mechanism was first proposed by Reeves, Fowler \& Hoyle (1970) and required
energies of 150 Mev or more; it is referred to as Galactic Cosmic Ray (GCR)
spallation.  More specific details are found in Meneguzzi, Audouze \& Reeves
(1971).

Inasmuch as these elements are so rare, they need to be observed in their
resonance lines.  However, for Li this is the Li I doublet at 6707.7 and
6707.9 \AA.  The ionization potential for Li I is 5.39 eV which means that for
almost all stars most Li is in the form of (unobservable) Li II with resonance
lines in the extreme ultraviolet.  It also means that even the 6707 \AA{}
resonance doublet of Li I can be weak in solar-type stars.  For Be I the
resonance line is in the near UV at 2349.6 \AA{} so it is the Be II resonance
doublet that we use at 3130.4 and 3131.1 \AA.  The ionization potential of Be
I is 9.32 so most of the Be is in the form of Be II in solar-like stars.  The
strongest B lines are in the satellite ultraviolet so the spectrograph needs
to be above the Earth's atmosphere.  Although most of the B is in the form of
B II in stars like the Sun, we observe it primarily as B I at 2497.7 \AA.
Observations have been made of the B II resonance at 1362.5 \AA{}
(e.g. Boesgaard \& Praderie 1981) and some have been made of B III at 2066
\AA{} (e.g. Mendel et al.~2006).

\section{OBSERVING THE LIGHT ELEMENTS}

\subsection{Lithium}
The easiest of the three light elements to observe is Li.  The Li I resonance
doublet is in the red region of the spectrum at 6707.76 and 6707.91 \AA{} and
relatively free of other spectral features.  Only the coolest stars have
blending lines of molecules such as TiO and CN.  A labeled example of the Li
spectral region in an F9 star in NGC 752 is shown in Figure 5 from Boesgaard
et al. (2022).  The strong neighboring lines are due to Fe I.  The Fe I
feature that is just shortward of the Li I doublet must be taken into account
in determining the Li abundance, especially in cooler stars where it grows
stronger.  The line list used for the Li I doublet should include hyperfine
structure and possibly the presence of the isotopic lines of $^6$Li and the Fe
I line at 6707.4 \AA.  Stellar rotation can cause broadening of the line
profile and needs to be taken into account, especially in rapid rotators.

Examples of spectrum synthesis fits covering the same 8 \AA{} segment of Li
are shown in Figure 6 for two Hyades stars.  The synthesis fit includes the
three prominent lines of Fe I; the weak blending Fe I line can be discerned in
the fit at 6707.4 \AA.  (This Figure is from Boesgaard et al.~2016.)

\begin{figure}[htb!]
\epsscale{0.45}
\plotone{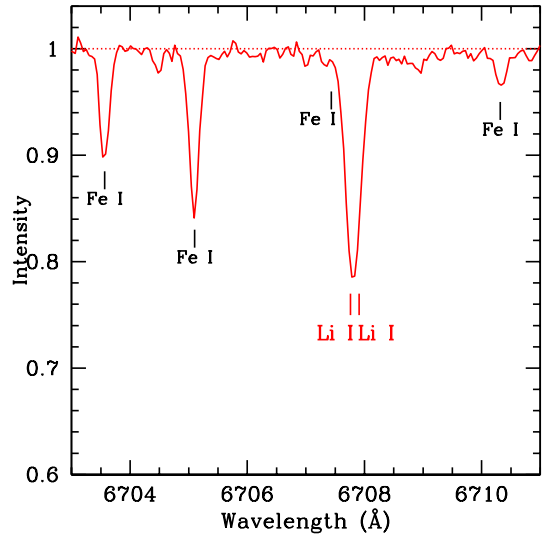} 
\caption{Eight \AA{} in the Li I spectral region showing the Li I doublet and
neighboring Fe I lines.  This is an F8 dwarf star, T = 6280 K, in NGC 752.
Similar to Figure 3 from Boesgaard, Lum, Chontos, et al.~2022).}
\end{figure}

\begin{figure}[htb!]
\epsscale{0.45}
\plotone{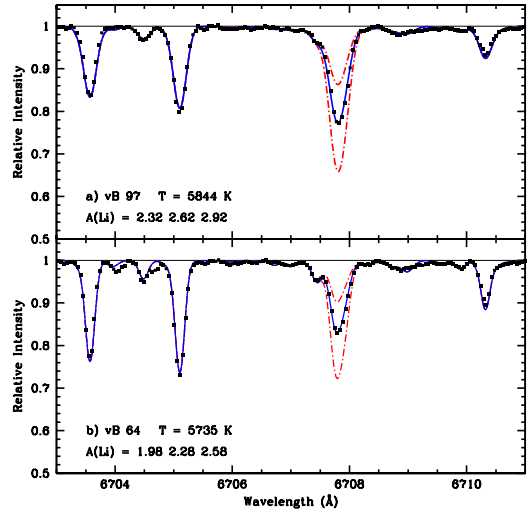} 
\caption{Examples of a spectrum synthesis fit in the Li region for two Hyades
stars.  The black dots are the observations and the solid blue line shows the
best fit Li abundance.  The red dashed-dotted lines show a factor of two more
and a factor of two less Li.  (Similar to Figure 9 from Boesgaard, Lum,
Deliyannis et al.~(2016).}
\end{figure}

\subsection{Beryllium}
Although Be II can be observed in stars from ground-based observatories, the
resonance doublet lines at 3130.422 and 3131.067 \AA{} are near the
atmospheric cutoff ($\sim$3000 \AA{}.)  The atmospheric transmission at those
wavelengths is poor and atmospheric refraction effects must be taken into
account during the observations.  In addition, this is a very crowded region
in the spectra of F-type and cooler stars.  Figure 7 shows an 8 \AA{} region
of spectra in the vicinity of the Be II lines; a few of the blending lines are
identified.  The line-crowding is evident in this section of spectrum,
especially in comparison to Figure 5 of Li I.  Both the Li figure and the Be
figure show a segment of 8 \AA.

\begin{figure}[htb!]
\epsscale{0.45}
\plotone{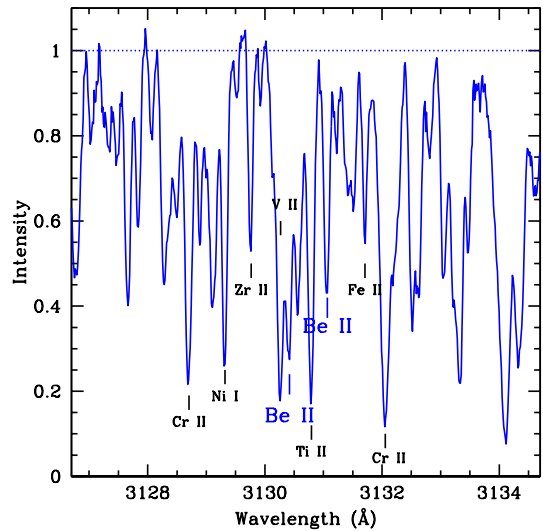} 
\caption{Eight \AA{} in the Be II spectral region showing the Be II doublet
and identifying some of the neighboring lines.  This is the same star as shown
for the Li region in Figure 5.  Unlike the Li I region, this spectral region
is very crowded with lines of atoms and molecules; only 8 \AA{} is shown.
Similar to Figure 2 from Boesgaard, Lum, Chontos et al.~(2022).}
\end{figure}

Examples of the synthesis of 2 \AA{} in the Be II region for two field stars
with similar temperatures are shown in Figure 8.  There are over 300 atomic
and molecular lines used in the spectrum synthesis spanning the 3 \AA{}
region.

Although for metal-poor stars most of the blending lines in the Be II
region are weaker, the lines due to the electronic transition lines of the OH
molecule are stronger and [O/Fe] is larger.  Those OH features must be taken
into account and the abundance of O needs to be determined through the
spectrum synthesis.  Examples of this process and results are shown in Figure
9 for a star with [Fe/H] = $-$1.0 and one with [Fe/H] = $-$2.8.  The
abundances of both Be and O can be determined for these metal-poor stars.
Three of the OH lines are shown in that region of the syntheses as labeled in
the figure caption.

\begin{figure}[htb!]
\epsscale{0.4}
\plotone{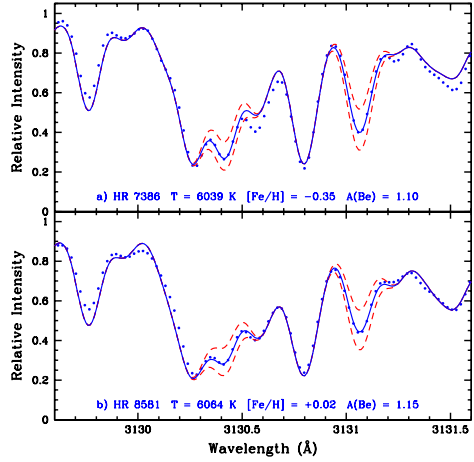}
\caption{Synthesis of 2 \AA{} in the Be II spectral region of two field stars.
The black dots are the observations and the blue line is the best synthetic
fit.  The dashed red lines are a factor of 2 more and a factor of 2 less Be.
From Figure 2 in Boesgaard, Armengaud, King et al.~(2004b).}
\end{figure}

\begin{figure}[htb!]
\epsscale{0.4}
\plotone{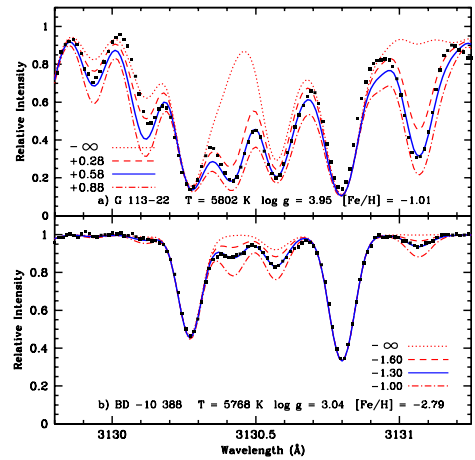} 
\caption{Synthesis of 1.45 \AA{} in the Be II region of two metal-poor stars
with similar temperatures but very different metallicities.  The black squares
are the observations and the solid blue lines are the best fits for Be and O.
Abundances of Be are shown as a factor of two greater - dot-dashed line - and
a factor of two lower - long-dash line and also a version with no Be at all as
the red dotted line.  Four values for the O abundance are shown differing by
0.2 are shown for the OH features at 3129.94, 3130.13, and 3130.57 \AA.  This
is from Figure 2 from Boesgaard, Rich, Levesque et al.~2011.}
\end{figure}

\clearpage
\subsection{Boron}
Of these three light elements B is most difficult to observe because the
resonance lines of the three common ions occur in spectral domains visible
only above Earth's atmosphere.  Most recent observations have been made with
spectrometers aboard the Hubble Space Telescope.  The feature in the spectrum
of B that is usually observed  is the resonance line of B I at $\lambda$2496.
Figure 10 shows a 2.6 \AA{} section of four field stars showing the position of
the B I line.  The stars have a range in temperature of almost 500 K.

\begin{figure}[htb!]
\epsscale{0.6}
\plotone{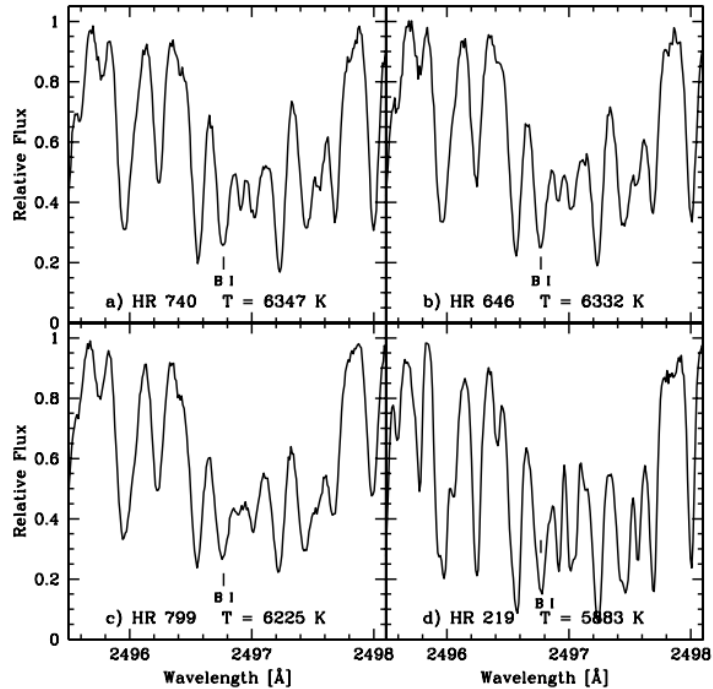} 
\caption{Examples of B spectra for four field stars at the specified
temperatures observed with STIS/HST at spectral resolution of 114,000.  The B
I line is indicated for each star.  (Figure from Boesgaard, McGrath, Lambert
\& Cunha 2004.) }
\end{figure}

The spectrum is very crowded which presents a challenge to extract a B
abundance even with sophisticated spectrum synthesis techniques and line
lists.  Examples of the synthesis fits are shown for two stars in Figure 11.

\begin{figure}[htb!]
\epsscale{0.45}
\plotone{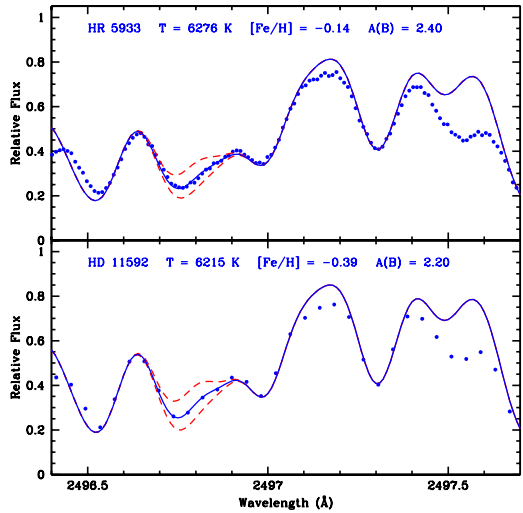} 
\caption{Spectrum synthesis of B I of HR 5933 and HD 11592.  The spectrum of
HR 5933 has a resolution of 114,000, while that of HD 11592 has a resolution
of 30,000.  The dots are the observed points and the solid line shows the best
fit.  The dashed lines show B abundances a factor of two higher and lower.
The two stars have similar temperatures, but HD 11592 has a lower metallicity
and a lower abundance of B.  (Figure from Boesgaard, McGrath, Lambert
\& Cunha 2004.)}
\end{figure}

\subsection{Effects of Stellar Rotation}

Any amount of spectral line broadening increases the difficulty of determining
accurate abundances. The largest amount is due to stellar rotation which is
measured as v sin i.  This issue is illustrated in Figure 12, left, with the
synthesis fits for two Hyades stars.  The observed spectra have high spectral
resolution and high signal-to-noise ratios so Li abundance determinations are
very good for both stars.  The difference in the line-broadening is obvious
with the sharp-lined star, vB 62, with v sin i = 4.8 km s$^{-1}$ and the star
with broader lines, vB 121, at v sin i = 15.9 km s$^{-1}$ (The v sin i values
are from Mermilliod, Mayor \& Udry (2009).

\begin{figure}[htb!]
\epsscale{1.1}
\plottwo{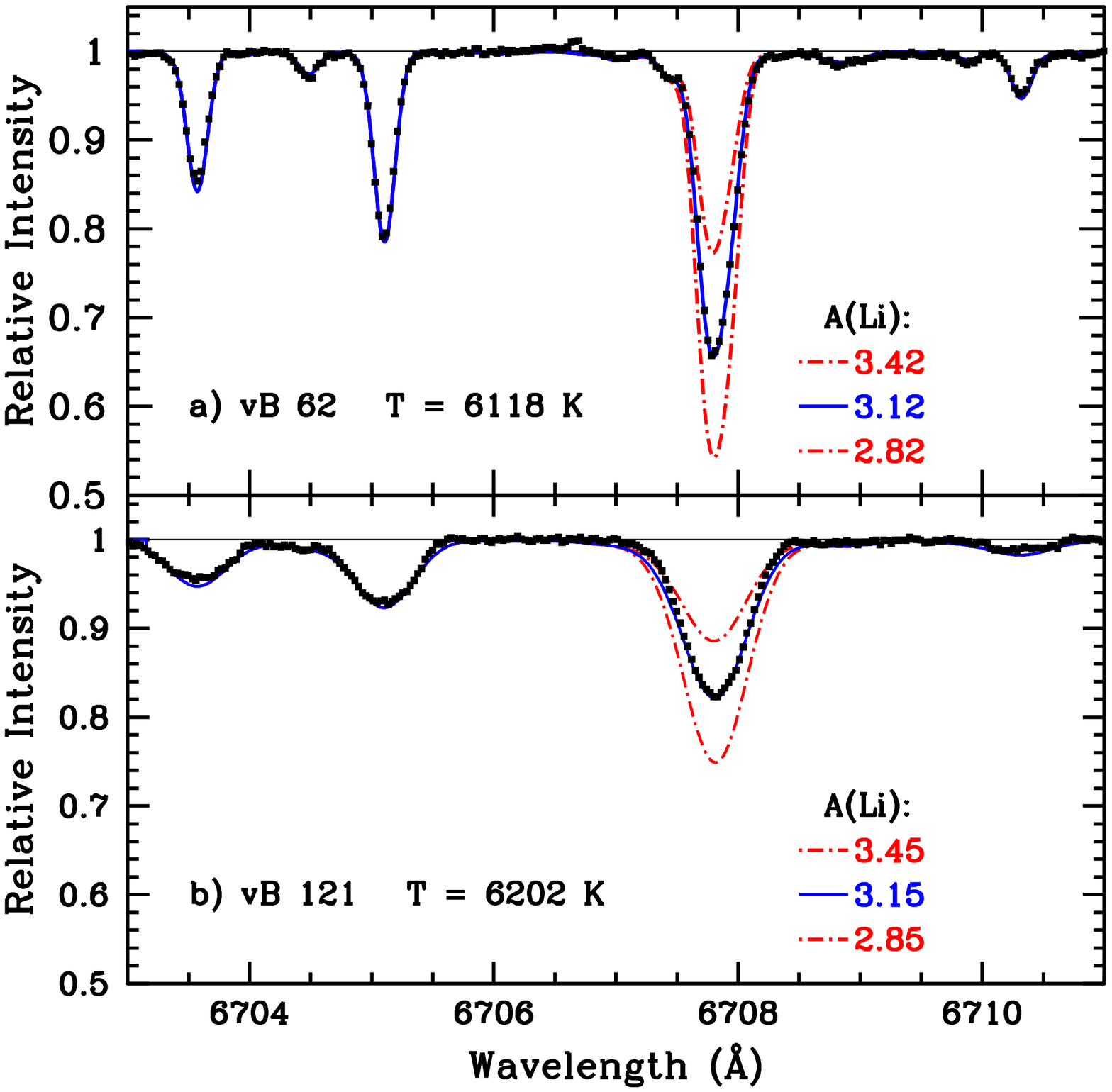}{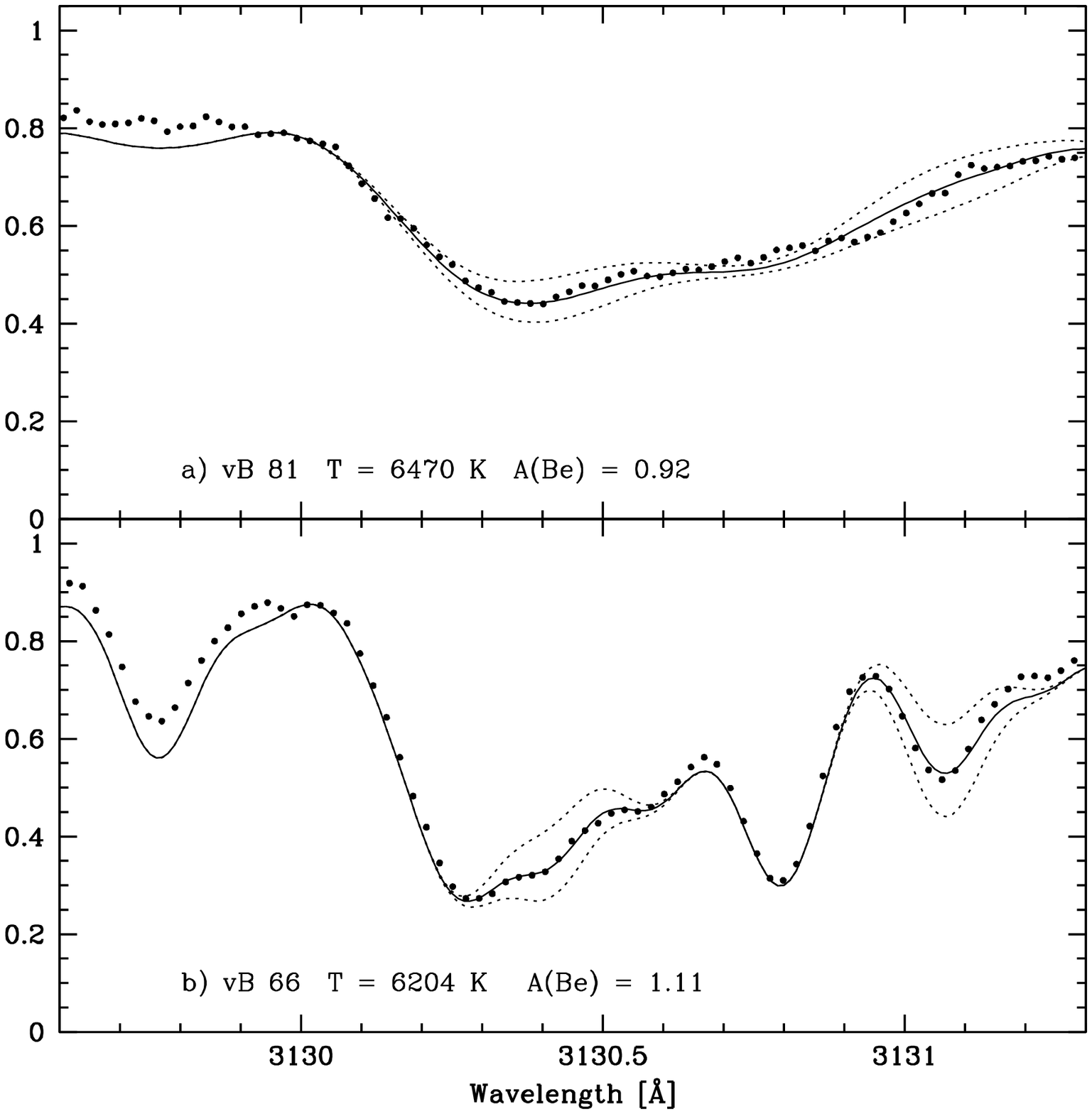}
\caption{Left: Spectrum syntheses for Li in two Hyades dwarf stars.  The upper
one, vB62, has v sin i = 4.8 km s$^{-1}$ while the lower on, vB 121 has v sin
i = 15.9 km s$^{-1}$ (Mermilliod, Mayor \& Udry (2009).  The best fit to the
observed (black) points is the solid blue line.  The red dash-dot lines are a
factor of two higher and lower in Li.  Right: Spectrum synthesis for Be in two
other Hyades dwarf stars.  The points are the observed spectra.  The syntheses
show a best fit (solid line) and a factor of two higher and lower than that
fit (dotted lines).  For vB 81 shown in the upper panel, the value of v sin i
is 24.5 km s$^{-1}$ and for vB 66 in the lower panel v sin i is 10.2 km
s$^{-1}$ (Mermilliod, Mayor \& Udry (2009).}
\end{figure}

The determination of reliable results for both Be and B is dramatically
impacted by the effects of stellar rotation.  The resonance lines of Be II and
B I occur in regions of the spectrum with many, many blending lines.  Those
lines all become blurred and the Be and B lines are less distinct.  This can
be seen in the eight \AA{} region of Be II in Figure 7 and the six \AA{}
region of B I in Figure 10.  So it is crucial to know the atomic properties of
those blending lines with precision.  Even so, extracting accurate values in
rotating stars is extremely challenging.  Figure 12, right, shows the
difficulty for Be with two Hyades stars.  For vb 81, with v sin i = 24.5 km
s$^{-1}$, the lines are very blended by rotation.  Even vB 66 with v sin i =
10.2 km s$^{-1}$, the line-broadening is troublesome.  (The Be synthesis fits
in Figure 8 are for stars with v sin i values of $\leq$6 km sec$^{-1}$.)

\section{GALACTIC PRODUCTION AND EVOLUTION OF THE LIGHT ELEMENTS}

\subsection{Lithium}

The present-day evidence of the Big Bang Li production was discovered in 1982
by Spite \& Spite (1982) who found nearly constant Li abundances in the
metal-poor stars they were studying and plotted them against temperature.
Rebolo et al.~(1988) displayed the Li abundances with [Fe/H] in many stars.
This showed not only the constancy of Li at A(Li) at $\sim$$-$2.2
below [Fe/H] of $-$1.0 as the result of that primordial Li production, but
also, at higher metallicities, both the galactic production of more Li and its
destruction in individual stars.  A(Li) = log N(Li)/N(H) + 12.00.

\begin{figure}[htb!]
\epsscale{0.7}
\plotone{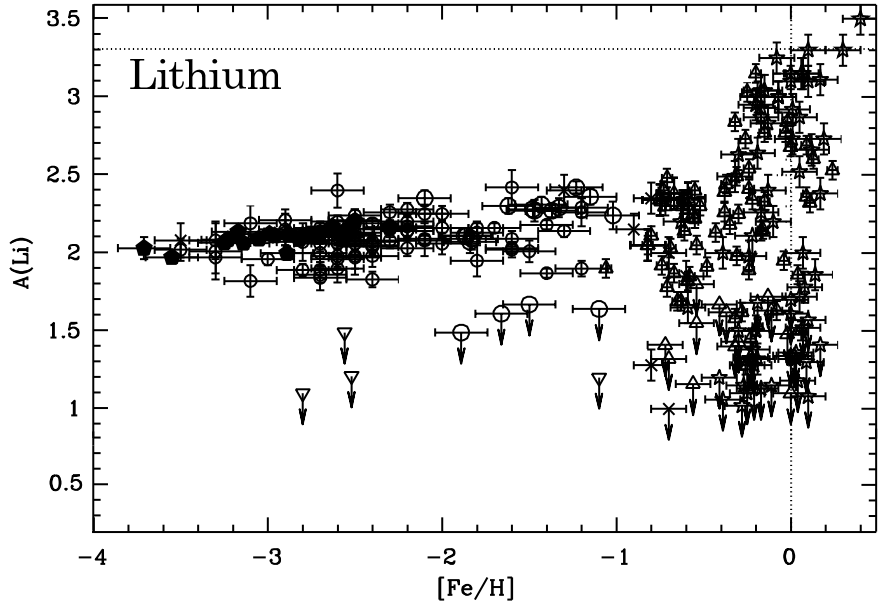} 
\caption{The relationship between the abundances of Li and Fe similar to
Figure 6 from Ryan, Kajino, Beers et al. (2001).  The ``Li plateau'' from Big
Band Li production in the metal-poor dwarf stars extends from [Fe/H] near $-$4
to $\sim$$-$1.0.  Some stars in the plateau region are clearly severely Li
depleted.  There is a strong rise in Li abundances seen at higher
metallicities due to overall galactic production of the light elements.  There
is also severe Li depletion in many of the more metal-rich stars resulting from
internal mixing processes and subsequent Li depletion.}
\end{figure}

Figure 13 is similar to Figure 6 in Ryan, Kajino, Beers et al.~(2001) and
shows Li abundances with [Fe/H] from $\sim$$-$4.0 to $\sim$0.0.  Almost all of
the stars with [Fe/H] below about $-$1.0 have the Li that was produced during
the Big Bang.  A few of metal-poor stars show Li depletions with only upper
limits on the Li abundances.  In this figure the metal-richer stars, ([Fe/H]
$\geq$$-$0.4), show both galactic enhancements from light element production
(e.g. GCR spallation) and stellar depletions of Li from internal mixing.
Additional Li results at even lower values of [Fe/H], down to $-$6, can be
found in Bonifacio et al.~(2018), Aguado et al.~(2019).

\subsection{Beryllium and Boron}

There is no expectation of primordial production of Be or B so no such
plateaus with [Fe/H] were expected nor found.  Both elements show a smooth
enhancement with increasing Fe.  This can be seen for Be in Figure 14 based on
work from Boesgaard et al.~(2011), supplemented by abundances from Boesgaard
\& Novicki (2005, 2006) and Boesgaard \& Hollek (2009) primarily from Be data
obtained with the Keck I high-resolution spectrograph.  This covers a range in
[Fe/H] of 4 orders of magnitude from $-$4.0 to 0.0 and shows a steady increase
in A(Be) from $-$2.0 to +1.4 with a slope of 0.86 $\pm$0.02.  

\begin{figure}[htb!]
\epsscale{0.5}
\plotone{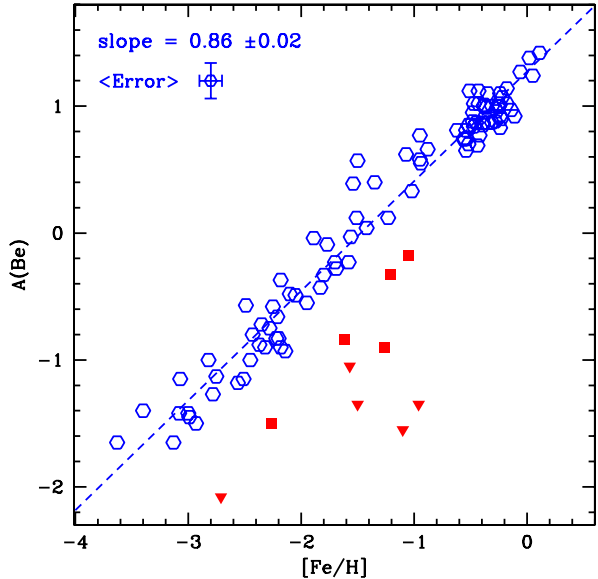} 
\caption{The counterpart diagram to Figure 13 for Be.  There is no evidence
for early production of Be in the Big Bang (the $^7$Be decays to $^7$Li), but
the uniform Galactic production of Be with metallicity/age can be seen all the
way to solar metallicity stars.  The Be-deficient stars - red squares and
triangles (upper limits) - are all ultra Li-deficient.}
\end{figure}

It can be seen in Figure 13  that there are some old, metal-poor stars with
large Li deficiencies, i.e. stars well below the Li-plateau level for stars
with the Big Bang Li.  Boesgaard (2007) studied Be in seven of the nine very
Li-deficient stars and found large Be-deficiences as well.  Predictions from
models with rotationally-induced mixing of Pinsoneault et al.~(1992) do not
deplete enough Be to account for the low levels of observed Be.  The lack of
both Li and Be could be attributed to mass transfer events or stellar mergers
in binary stars which would cause mixing of material to deeper layers in the
star.  Such deep mixing would result in destruction of more Li and Be.

Figure 15 shows the B abundances with [Fe/H] in Duncan et al.~(1997) and
Duncan et al.~(1998) in metal-poor stars.  These observations came from the
Goddard High-Resolution Spectrograph on the Hubble Space Telescope.  Their
sample of 12 stars covers a range of 3 orders of magnitude in [Fe/H].  The B
abundance, as corrected for non-local thermodynamic equilibrium effects (nLTE)
, also increases over 3 orders of magnitude.  The slope of the relation is
0.96 $\pm$0.07.

\begin{figure}[htb!]
\epsscale{0.55}
\plotone{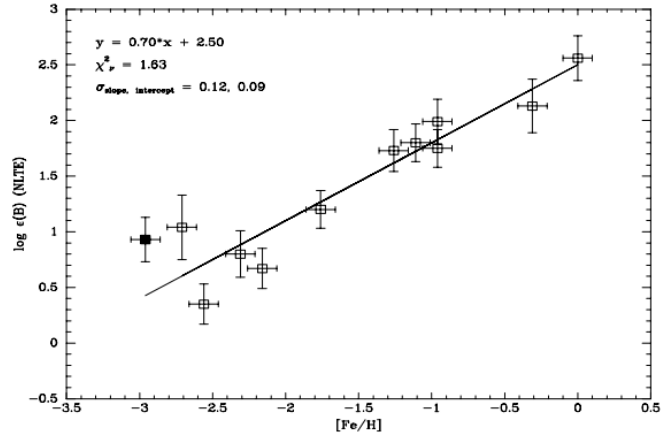} 
\caption{The relationship between the nLTE B abundance and metal content from
Duncan, Primas, Rebull et al.~(1997) and Duncan, Rebull, Primas et al.~(1998).
The slope is 0.71.}
\end{figure}

For both Be and B the abundances are done with spectrum synthesis fits.  For
both elements this is easier in metal-poor stars because the blending lines
are all much weaker than in solar metallicity stars.  As mentioned above for
Be it is necessary to evaluate the O abundance also due to the larger ratio
[O/Fe].  (See Figure 9.)

\subsection{Abundance Evolution with Time}

Boesgaard et al.~(2011) determined Be abundances along with O, Ti, Mg and Fe
in a sample of 117 metal-poor stars covering more than three orders of
magnitude in [Fe/H].  They found a steady increase in A(Be) with [O/H] as
shown in their Figures 11 and 12.  Inasmuch as Li, Be, and B can be produced
by energetic cosmic-ray spallation on elements like O in the interstellar
medium, this connection was to be expected.  Their Figure 13 also showed a
rather monotonic increase in [O/H] with [Fe/H] with a slope of 0.75 $\pm$0.03
over more than two and a half orders of magnitude in [O/H].  The increases in
al of those abundances is attributable to the various processes of galactic
chemical evolution.

In addition, their stars could be classified by their kinematics -- galactic
rotation velocity and apogalactic distance -- into components of the Galaxy.
There were equal numbers from the dissipative collapse population and from an
accretion-process population.  But the relationship between A(Be) and [Fe/H]
and between A(Be) and [O/H] were different for those two populations.  For the
dissipative group the slopes were near 1: 0.94 $\pm$0.04 for Be with Fe and
1.13 $\pm$0.08 for Be with O.  For the accretive population the slope between
A(Be) and [Fe/H] was 0.68 $\pm$0.04.  For A(Be) and [O/H] it was
0.76$\pm$0.06.  The accretive stars demonstrated a slower increase in Be
relative to both Fe and O than the classical disk and halo stars.

\section{STELLAR INTERIORS and LIGHT ELEMENT DESTRUCTION}

\subsection{Open Clusters}
Wallerstein, Herbig \& Conti (1965) studied Li in the Hyades with spectra from
photographic plates.  They measured line strengths of Li I at 6707 \AA{} and
7-12 lines of Ca I and determined the ratio of [Li/Ca] for 23 main sequence
stars.  They found a decline with increasing B-V, i.e. decreasing temperature.
For most of the hotter stars they could find only upper limits on the Li
abundances because the stars were rotating rapidly and the lines were too
broad for reliable line measurements.  The decline in Li was further defined
by Zappala (1972) with Li determinations in 41 Hyades FGK stars.  With high
quality reticon spectra from the CFHT coud\'e spectrograph of 12 Hyades stars
cooler than 6000 K, Cayrel et al.~(1984) determined that the decline with
temperature was even steeper than found in those prior studies.

Then Boesgaard \& Tripicco (1986) obtained high signal-to-noise spectra with a
reticon detector at the CFHT coud/'e spectrograph and with a CCD detector at
the UH 2.24-m telescope coud/'e spectrograph of Hyades stars in the
temperature range 6000 - 7100 K.  They found a dramatic drop in Li abundances
in stars near 6500 K with A(Li) more than 2 orders of magnitude lower than in
stars 300 K hotter or cooler.  Figure 16, left, shows the modern version of Li
in the Hyades with the F dwarf Li dip and the G dwarf Li fall off.  (Data from
Boesgaard, Lum, Deliyannis et al.~2016.)

\begin{figure}[b!]
\epsscale{0.8}
\plottwo{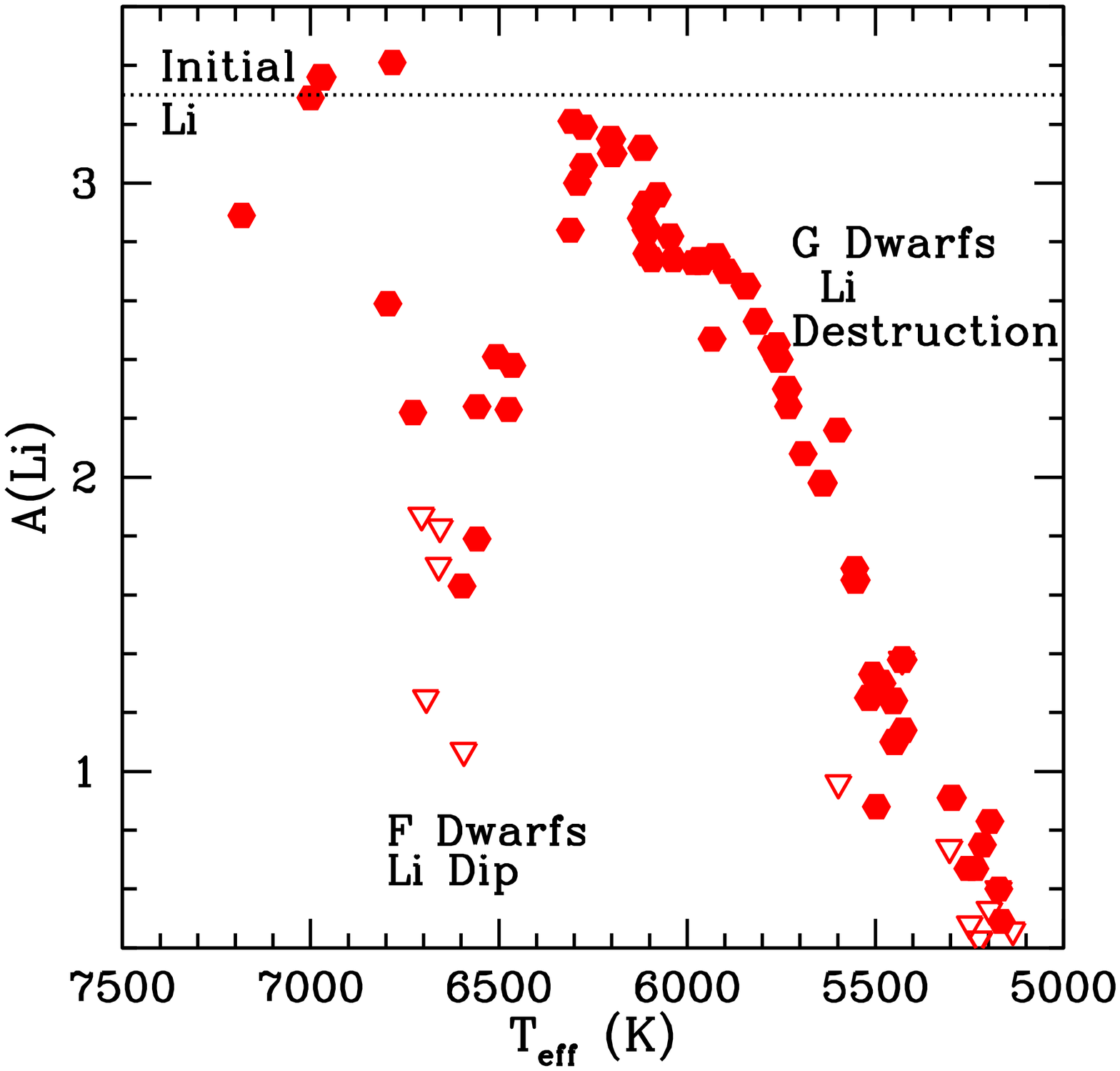}{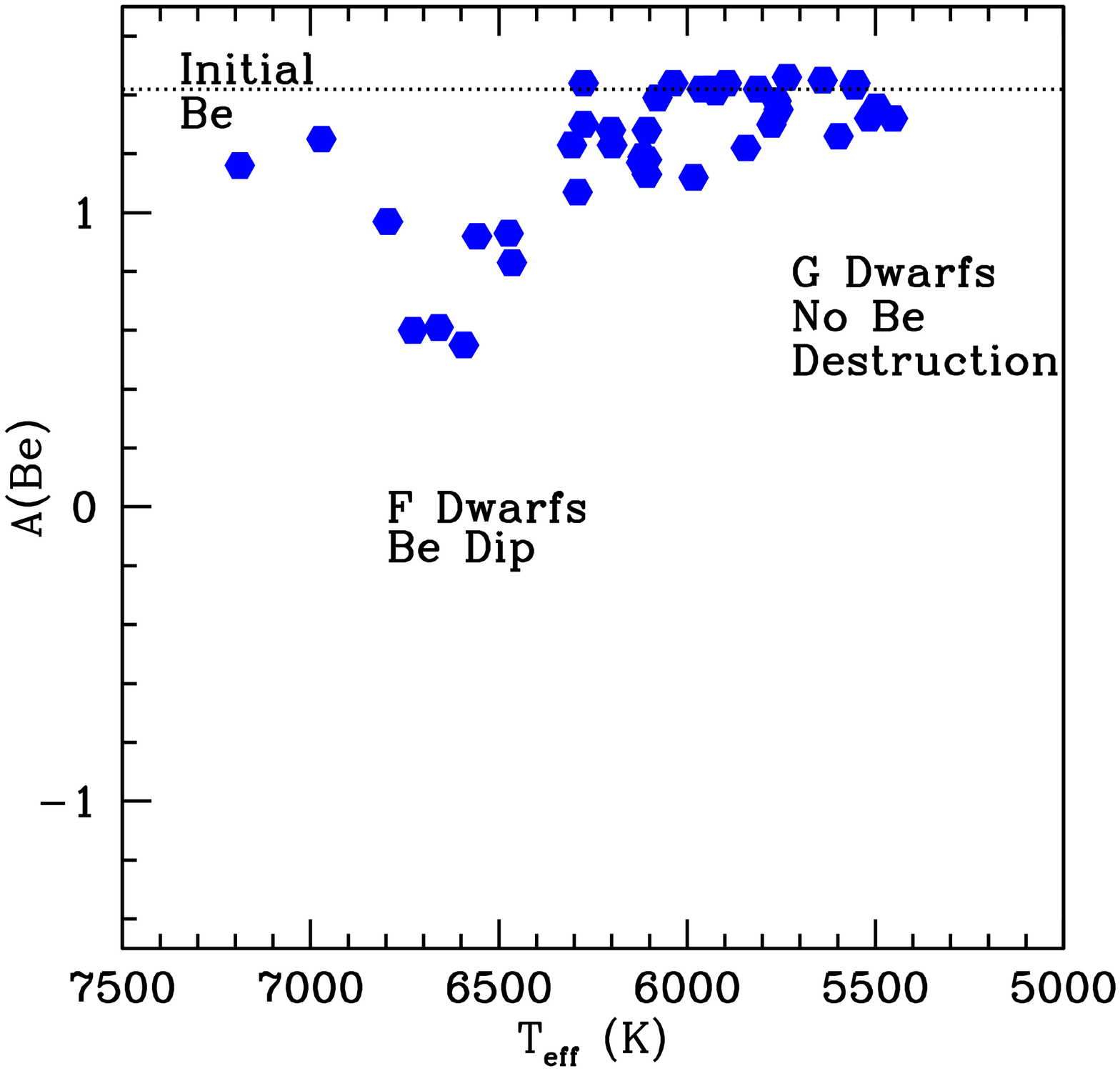} 
\caption{Left: The distribution of Li abundance with temperature in the
Hyades showing the deep Li depletions in the narrow temperature range centered
near 6600 K and the dramatic decline in Li with decreasing temperature for
stars cooler than $\sim$6100 K.  Right: The distribution of Be abundance in
the Hyades revealing a much shallower Be depletion in the mid-F dwarf
stars and no change in the Be abundance for the cooler stars.  Data are from
Boesgaard, Lum, Deliyannis et al.~(2016).}
\end{figure}

This result prompted both theoretical explanations of the Li dip and more
observations of Li in this temperature regime in other open clusters.  The
most interesting additional cluster to be observed was the (younger) Pleiades
which had little or no Li dip (see below).

A first attempt to determine Be in the Hyades stars was done by Boesgaard \&
Budge (1989) for eight stars.  Later observations were made of Be in 34 Hyades
stars with HIRES on Keck I by Boesgaard \& King (2002).  They discovered
evidence for a Be dip that was similar but not as deep as the Li dip.  They
also found that there was no decline in Be in the cooler stars as was found
for Li in the Hyades.  The modern version of this result is shown in Figure
16, right.  The Be observations were well-matched by models with
rotationally-induced mixing below the convection zone by Deliyannis \&
Pinsonneault (1997).  Figure 10 in Boesgaard \& King (2002) shows the Hyades
Be results with effective temperature and those model fits with initial
rotation values of 10 and 30 km s$^{-1}$.  Their Figure 11 shows the same
results for Li.

The observational results for for both Li and Be in the Hyades are shown
together as a function of stellar surface temperature in Figure
17, left, on the same scale and normalized to their respective solar system
abundances.

Observations of both Li $\underline{and}$ Be in 24 field stars with
temperatures between 5700 and 6700 K led Deliyannis et al.~(1998) to the
finding that Li and Be were rather tightly correlated with a slope of about
0.4.  This enabled them to exclude some of the explanations that had been
proposed for the Li (only) depletions.  For example, with mass loss accounting
for the Li depletion, all the Li would be lost before any effect would be seen
on Be.  With microscopic diffusion, which would affect the two elements
similarly, the slope would be close to 1.  The models of Deliyannis \&
Pinsonneault (1993, 1997) and Charbonnel et al.~(1994) involved slow mixing
caused by the effects of stellar rotation and provided an excellent fit to the
data.  This showed the importance of having abundances of two (or more) of the
light elements for the interpretation of the depletions of the light elements.
It demonstrated that the depletions were occurring simultaneously.

An important insight about this Li dip came from the study of Li in the young
Pleiades by Pilachowski, Booth \& Hobbs (1987) and Boesgaard, Budge \& Ramsey
(1988) who also included stars in the young $\alpha$ Per group.  There was no
Li dip as found in the mid-F dwarfs in the Hyades.  This showed that the
mechanism which caused the large depletion seen in the Hyades was a
main-sequence phenomenon and not a pre-main sequence occurrence.  

If there were no Li depletion in the Pleiades, no Be depletion was to be
expected.  Boesgaard, Armengaud \& King (2003a) studied Be in the young
clusters Pleiades and $\alpha$ Per.  Figure 17, right, reveals the combined Li
and Be results for the young Pleiades, again scaled and normalized as Figure
17, left, along with the fit through the Hyades Li data.  With an age of
$\sim$650 Myr the Hyades exhibits a pronounced dip in Li and a well-defined
dip in Be can be seen.  But there is no well-defined dip in either element in
the stars of the younger Pleiades at $\sim$130 Myr, a cluster one-fifth the
age of the Hyades.

\begin{figure}[htb!]
\plottwo{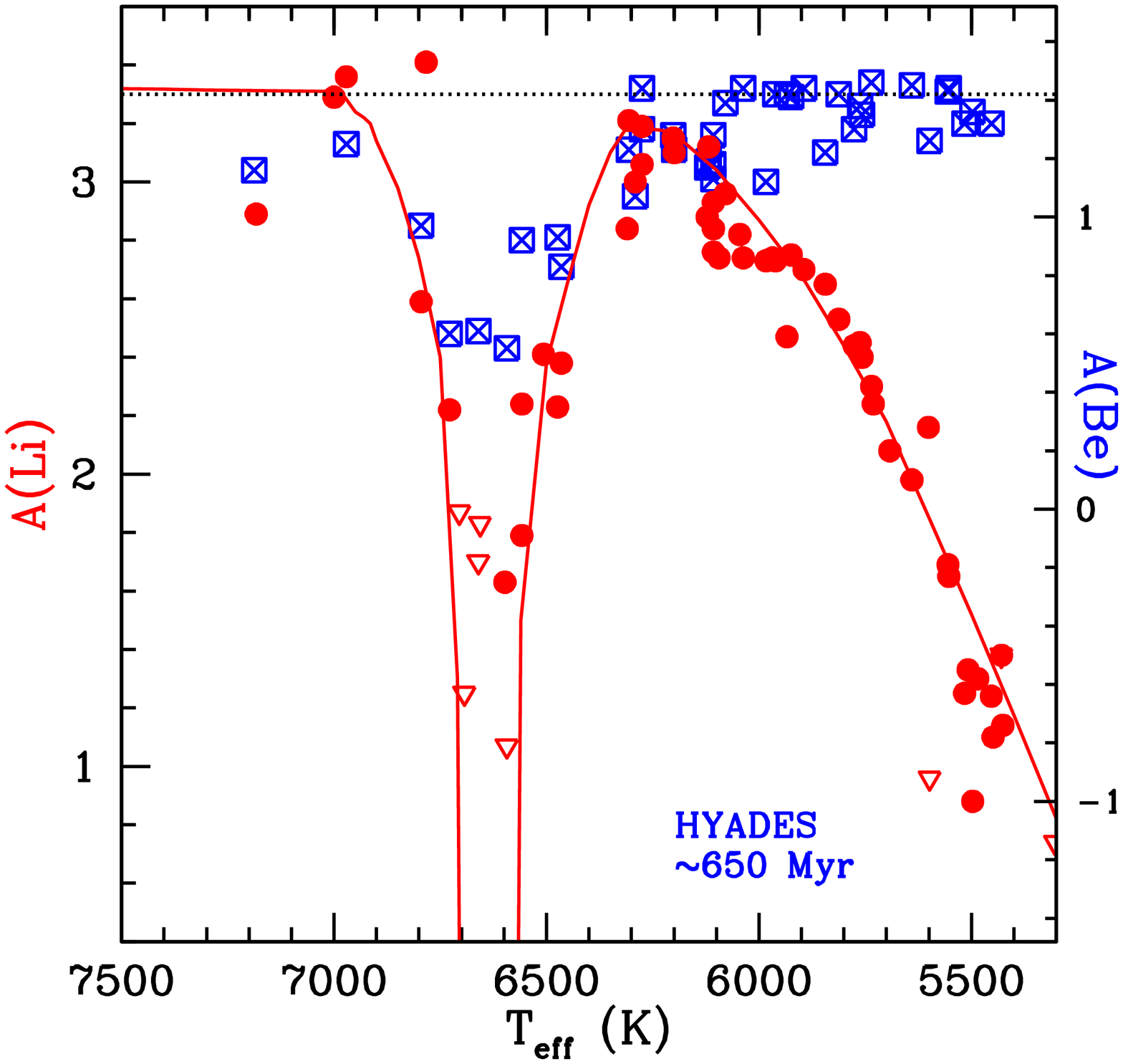}{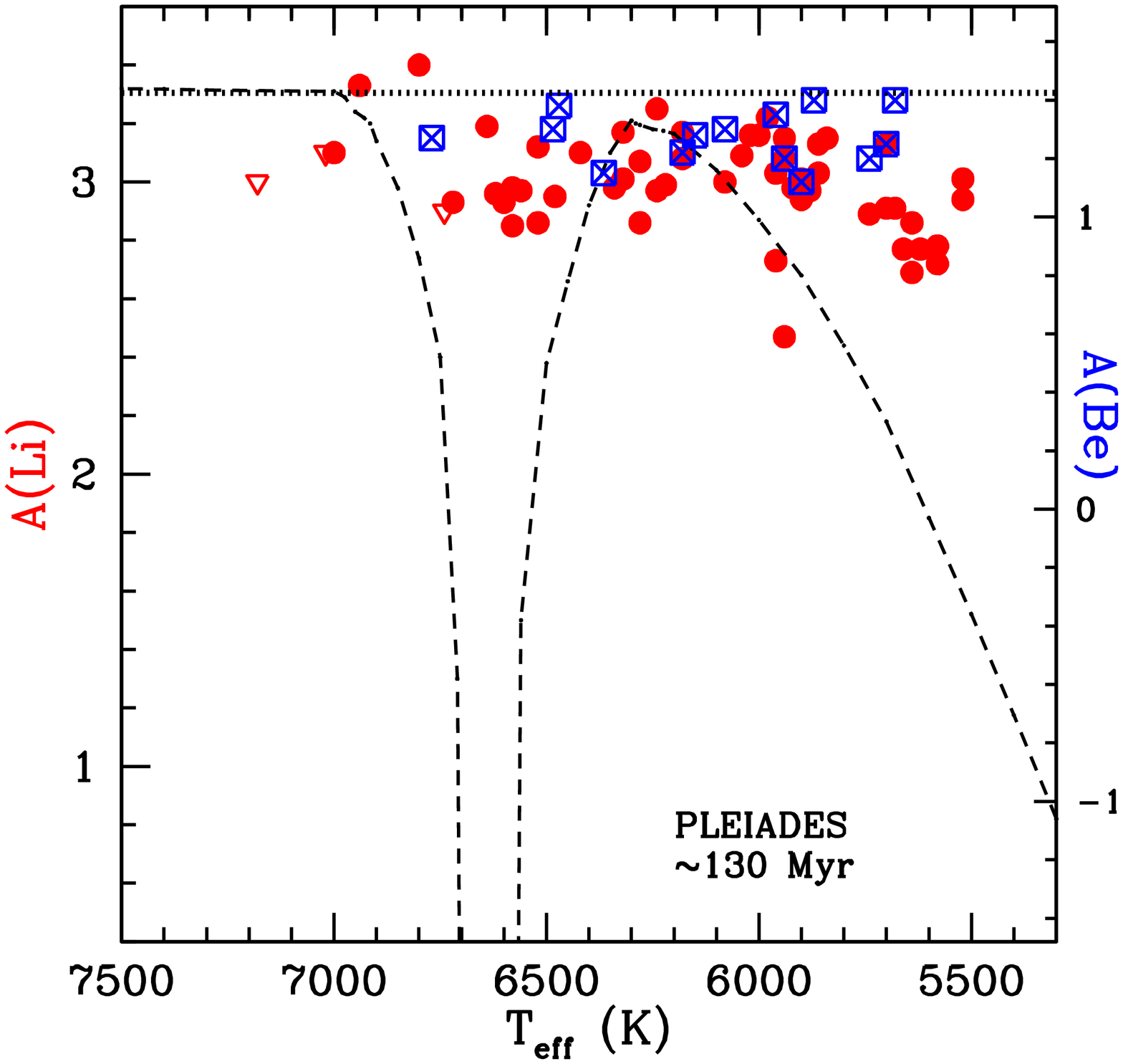}
\epsscale{0.9}
\caption{Left: Abundances of Li and Be on the same scale as a function of
stellar temperature for the Hyades, age $\sim$650 Myr.  There is some Be
depletion in the 6600 K region, but not nearly as much as found for Li.  There
are no upper limits in the Be abundances.  The steep decline in Li from T
$\sim$6200 to 5300 K is not seen for Be.  Right: Abundances of Li and Be on
the same scale as a function of stellar temperature for the younger Pleiades
cluster at age $\sim$130 Myr.  There is little if any depletion of Li or Be in
the younger Pleiades.}
\end{figure}

This extra internal-mixing effect - caused by rotation and spindown - showed
that it was very important in depleting surface Li, but less effective for Be
in the mid-F dwarf stars .  An important next step was to determine the effect
of this internal mixing on B.  As mentioned above, B requires an even higher
temperature, $\sim$5 x 10$^6$ K, for thermonuclear reactions to destroy it.
It is also more difficult to observe, but ultimately we were able to use the
Hubble Space Telescope with its high resolution spectrograph to observe B in a
few Hyades stars.  The results for the Hyades can be seen in Figures 18, left,
for B and Li and Figure 18, right, for B and Be.  The two stars that are in
the Li-Be dip in the Hyades are depleted relative to the three on either side
of the dip.  The dotted lines connect those B points to their respective Li
and Be points.  They indicate that there is a measurable depletion on B also.
 
\begin{figure}[htb!]
\epsscale{0.9}
\plottwo{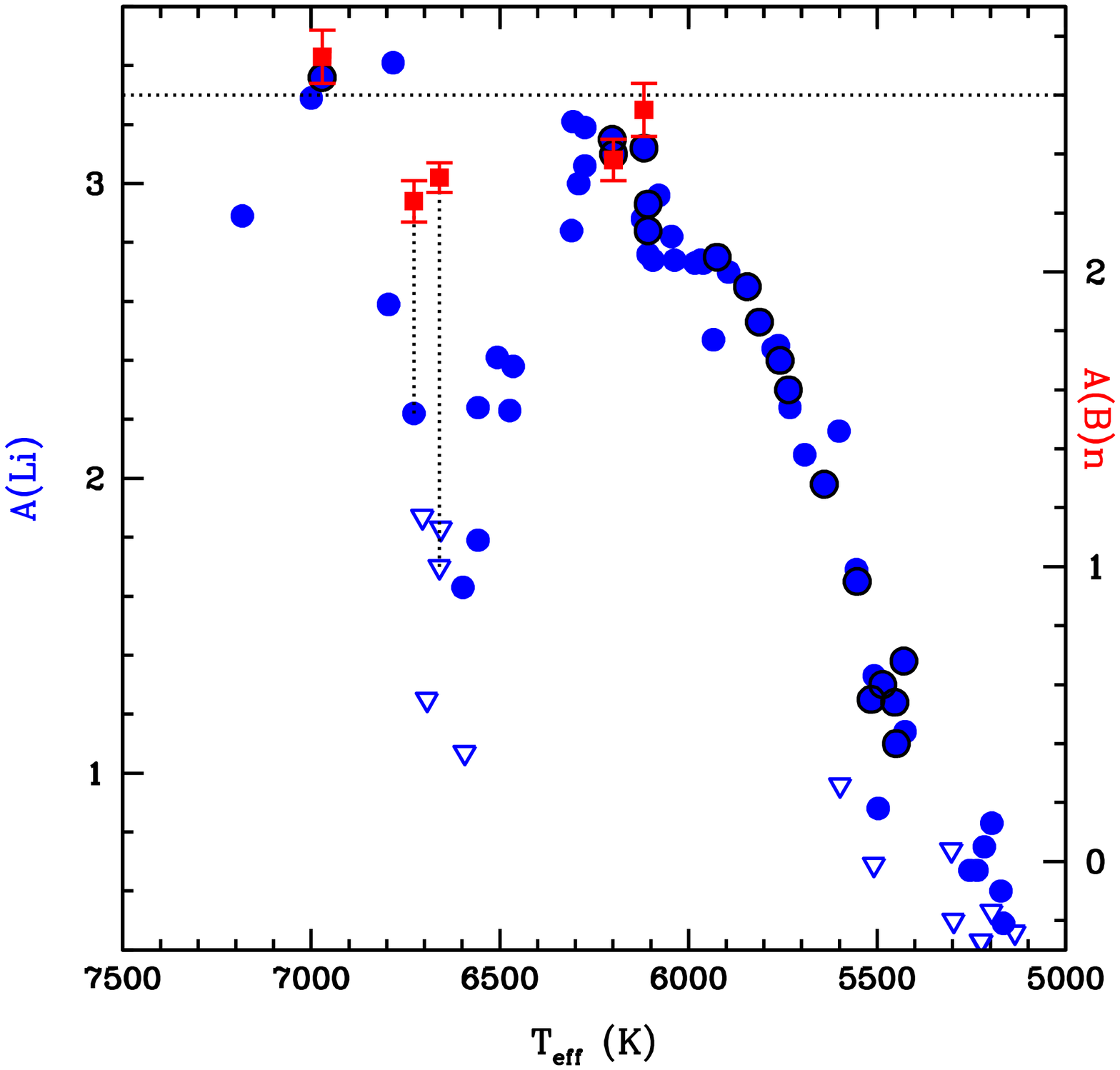}{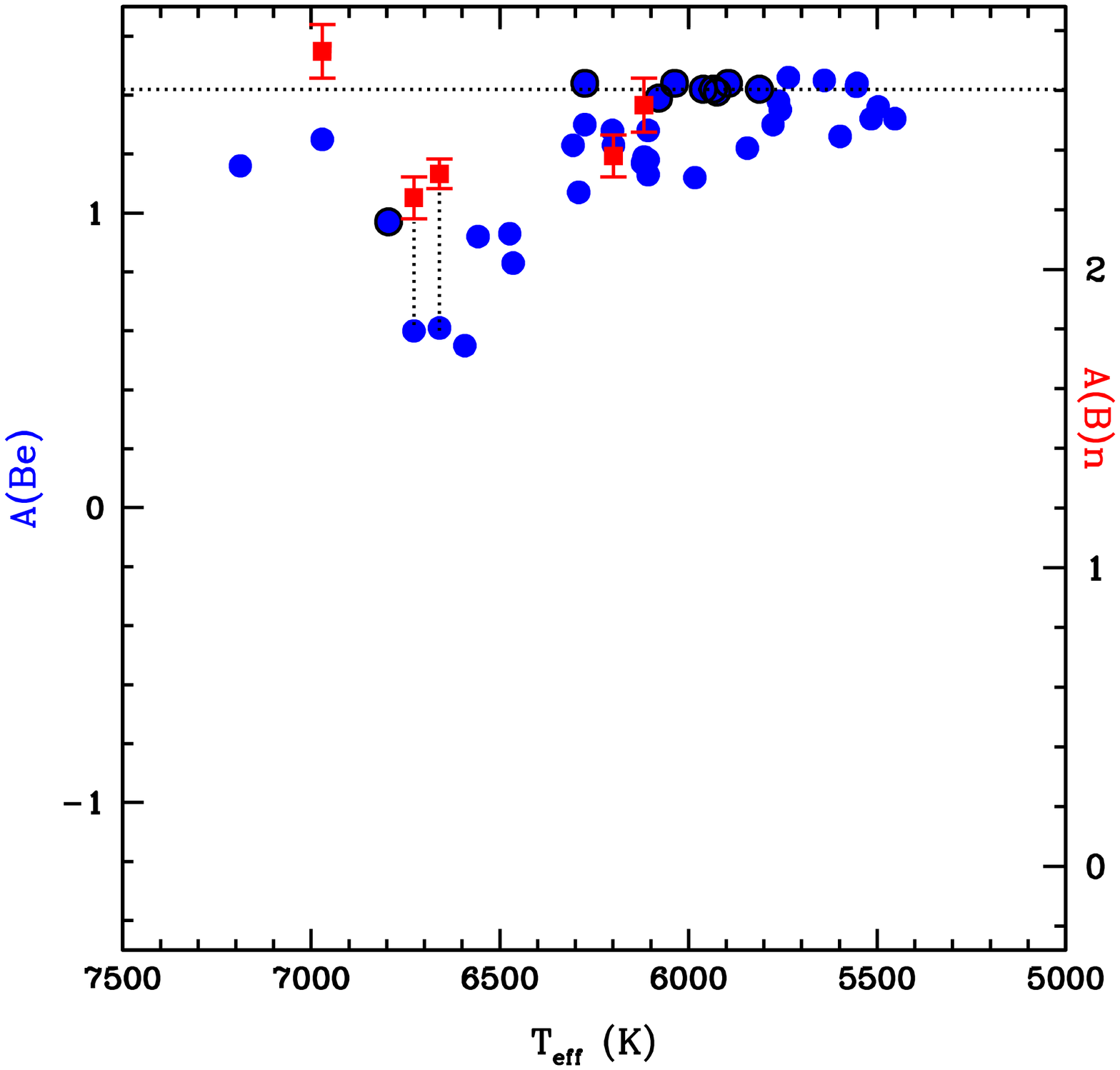}
\caption{Abundances of Li and B on the same scale as a function of stellar
temperature for the Hyades.  The two Li-dip stars observed for B are depleted
while the three stars outside the Li dip are not.  The dotted lines connect
the Li and B results for those two stars.  Right: The same figure with Be and
B results on the same scale.  Again, the two Li-Be dip stars are seen to be B
depleted; the results for those stars are connected by dotted lines.}
\end{figure}

A graphical depiction of a standard solar model is shown in Figure 19.  There
is an outer convection zone which is rather shallow in the Sun.  The nuclei of
atoms of Li have all been destroyed inside the red ring which encloses
temperatures higher than $\sim$2.5 x 10$^6$ K.  Atoms of Be are destroyed
inside the blue ring at $\sim$3.5 x 10$^6$ K and those of B inside the purple
ring at $\sim$5 x 10$^6$ K .  The surface shell which contains Li is very
small while the shell with Be extends deeper in the star to higher
temperatures and the shell with B is considerably larger.  That is the
reservoir with Li is smaller than that of Be and B and that of Be, while
larger than the reservoir for Li is smaller than that for B.  This diagram
illustrates two of the basic issues.  1) With the deepening of the surface
convection zone in main-sequence stars with decreasing surface temperatures,
the amount of Li on the surface will decrease as Li atoms are mixed to deeper
layers by convection.  However, convection alone is not sufficient to deplete
Li in the Sun.  Additional mixing mechanisms must be included in the standard
models.  2) The dip in the abundances of Li, Be, and B near 6600 K can not be
understood by standard models either, but the relative size of the depletions
there are determined by the susceptibility to thermonuclear destructive
reactions of each of the three elements.

\begin{figure}[htb!]
\epsscale{0.6}
\plotone{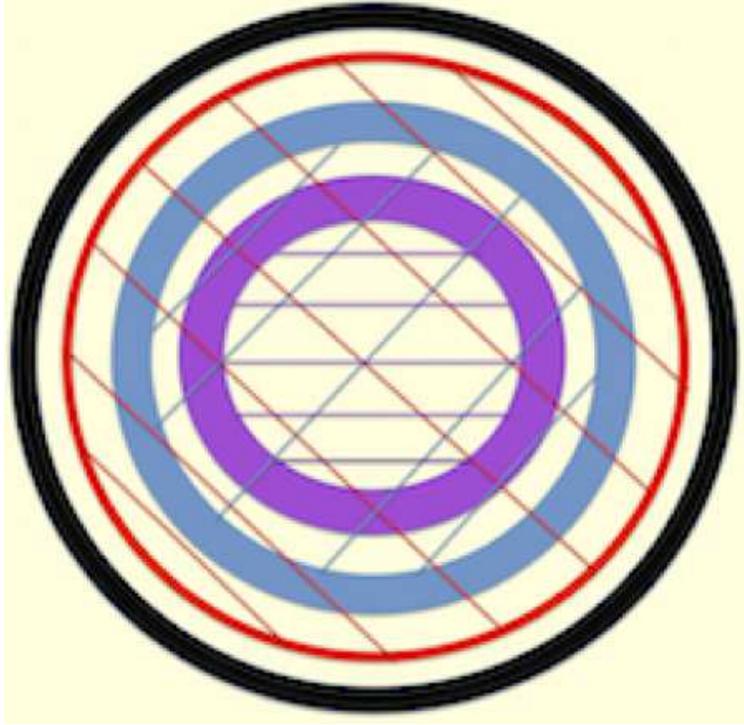}
\caption{A schematic image of a solar mass star.  Lithium nuclei are destroyed
inside the red circle with red diagonal lines, where the temperature is hotter
than $\sim$2.5 x 10$^6$ K.  Beryllium nuclei are destroyed inside the blue
circle with blue diagonal lines at $\sim$3.5 x 10$^6$ K and B nuclei are
destroyed inside the purple circle with purple horizontal lines at $\sim$5 x
10$^6$ K.  The black outer circle depicts the surface convection zone in the
Sun which does not penetrate deeply enough to mix the surface layers to
temperatures where the light elements would be destroyed.}
\end{figure}

\subsection{Correlation of Light Element Depletions}

Further observations of both Li and Be in main-sequence cluster stars and in
field stars showed how well-correlated the abundances of these two elements
are.  We studied Be in the young clusters Pleiades and $\alpha$ Per
(Boesgaard, Armengaud \& King (2003a), in the older Coma cluster and UMA
Moving Group (Boesgaard, Armengaud \& King (2003b), and in Praesepe and other
clusters (Boesgaard, Armengaud \& King (2004a) and put it together with field
stars in Boesgaard, Armengaud \& King (2004b).  These results provided a much
larger sample in which to investigate the correlation of Li and Be.  This can
be seen here in Figure 20, left, for stars on the cool side of the Li-Be dip,
T = 6000 - 6650 K.  The relationship has a slope of 0.44 $\pm$0.05.  This is a
remarkable relationship covering a range in Li of 400 times and Be of more
than 10 times for a large spread in age.

Both Be and B have been observed in field stars by Boesgaard et al.~(2005) and
in those five Hyades stars discussed above (Boesgaard et al.~(2016).  These
two elements are also well-correlated.  Figure 20, right, shows the Be-B
relationship for stars on the cool side of the Li dip T = 6100 - 6739 K) from
four Hyades stars and 14 field stars with detectable Be and B.  This
relationship has a slope of 0.22 $\pm$0.05.  The slope is shallower than that
for the Li-Be correlation (0.22 vs.~0.44) and extends over a range of 100 in
A(Be) and a range of 4 in A(B).  Models of stars with mixing induced by
rotation match the Be to B relation as discussed in Boesgaard et al.~(2016)
and seen in their Figure 15.

These observations show that Li declines more rapidly than Be.  They show that
Be declines more rapidly than B.  But we cannot compare Li and B because there
is no Li left when B starts to decline.

\subsection{Light Element Dilution}
 
In evolved stars the light elements can be diluted as well as depleted.  When
leaving the main sequence normal FGK stars expand and their outer convection
zones grow.  As mentioned above, each of these light elements has its own
outer shell of material which contains its ions.  As a star starts to expand
to become a red giant, this outer shell also expands.  The matter becomes
diluted by material containing no Li, Be, or B from deeper layers.  The effect
is first noticeable with Li as it is only in the outermost shell.  Then the
dilution will begin to affect Be and dilute its surface content.  Finally B,
with the deepest reservoir containing B ions, will become diluted also.  A
study of Li and Be in subgiants in the old open cluster, M 67, at 3.9 Gyr
shows the effects of both depletion and dilution of both light elements
(Boesgaard, Lum \& Deliyannis 2020).

\begin{figure}[htb!]
\epsscale{1.1}
\plottwo{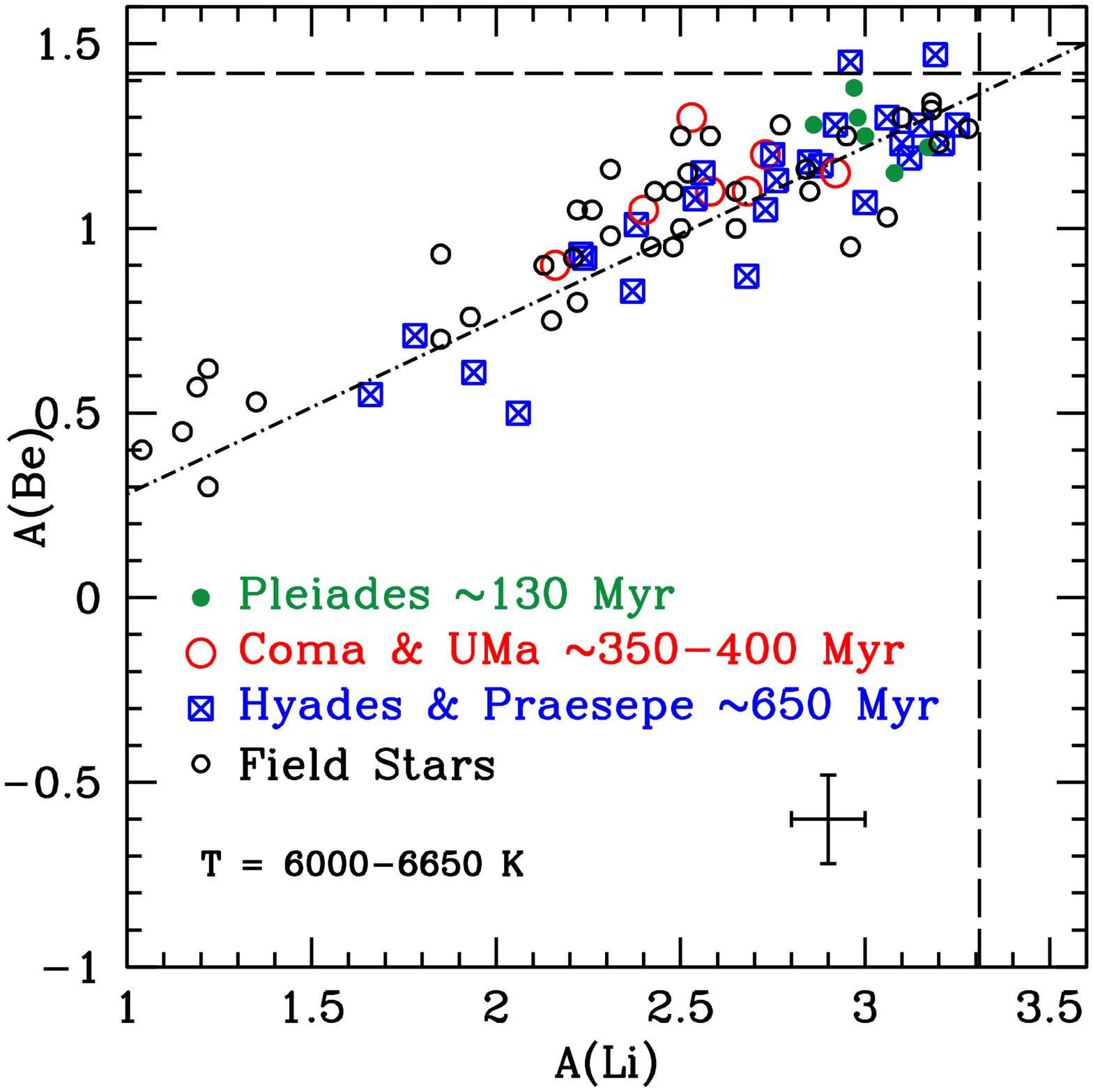}{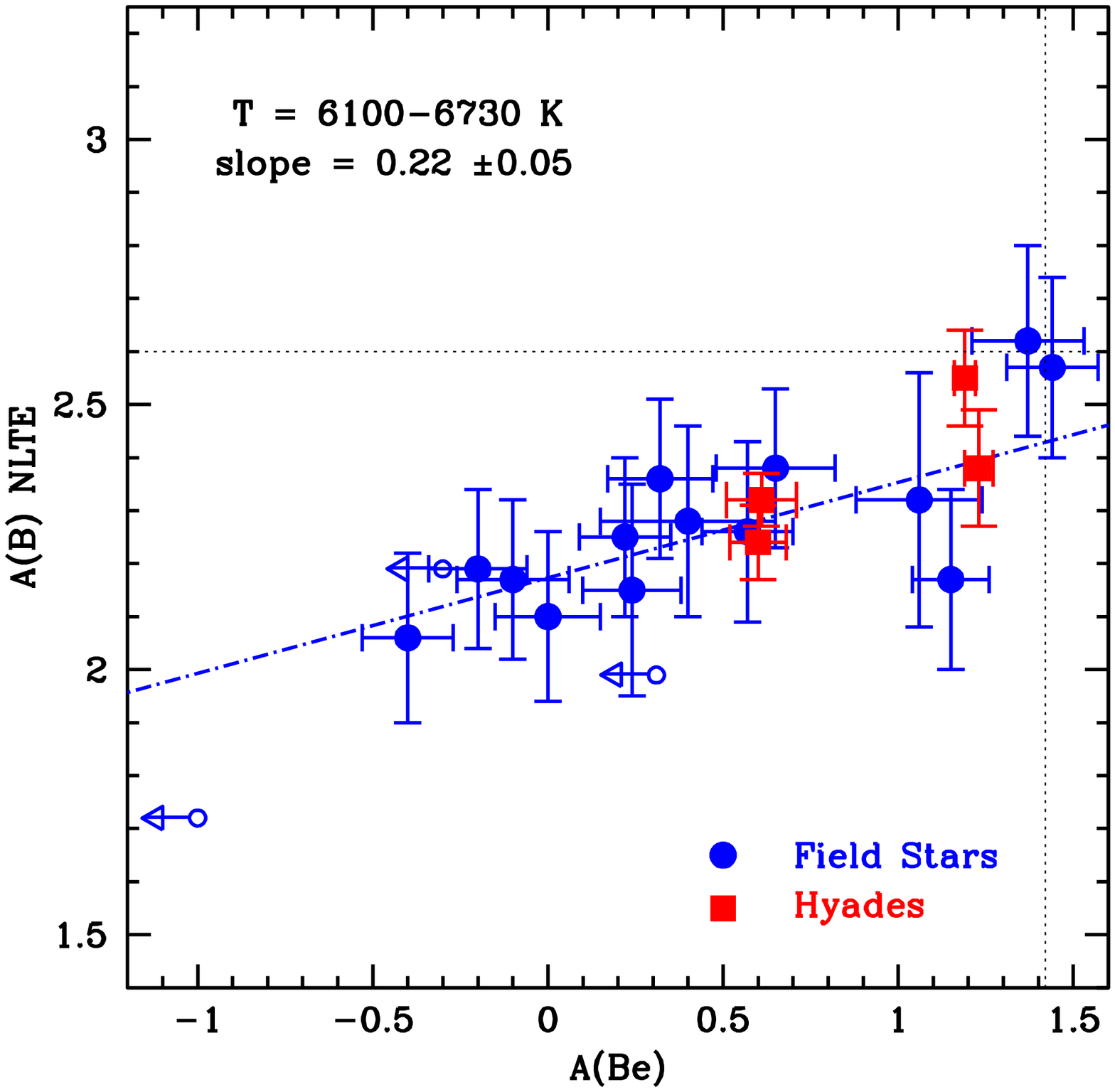}
\caption{Left: Li and Be in main-sequence dwarfs in cluster and field stars on
the cool side of the Li-Be dip region.  Both Li and Be depletions have
occurred, but Li is more depleted than Be.  The slope of this relationship is
0.44 $\pm$0.05.  The relationship spans over 200 times in Li abundance.
Right: Be and B in field and cluster dwarf stars on the cool side of the
Li-Be-B dip.  This figure is from Boesgaard, Lum, Deliyannis et al.~(2019),
Figure 14.  The slope is less steep than that for Li vs.~Be at 0.22 $\pm$0.05.
The depletion covers a range of $\sim$100 in Be abundance.  There is no
comparable diagram of Li and B because all the Li is destroyed before there is
measurable decline in B.}
\end{figure}

\section{SOME CONCLUSIONS}

In spite of their very low cosmic abundances, this trio of light
elements has produced some profound and interesting insights into stellar
structure and evolution.  The easiest to observe is Li which also provides
information about Big Bang nucleosynthesis.

The low abundance, however, means that observations must be made of the
strongest spectral features: the resonance lines.  For Li this is the Li I
6707 \AA{} doublet which is in an easily observable spectral region.  Due to
the low excitation potential of Li I, 5.39 eV, for most stars Li is in the
form of Li II, however.  Beryllium observations are made of the resonance
doublet of Be II at 3130 \AA{}, close to the atmospheric cutoff and in a
spectral region full of blending lines.  The resonance lines of B I, B II, and
B III are all in the ultraviolet spectral region observable by satellite only.

Most very low-metal stars show a plateau or a maximum in Li near A(Li) of 2.2
that is a product of Big Bang nucleosynthesis.  There are some, but not many,
stars at those low metallicities which show no Li feature.  Those
ultra-Li-deficient stars that have been studied for Be turn out to have Be
deficiences as well.  This dual deficiency could result from stellar mergers
or mass transfer events (see Boesgaard 2007).

Only in stars with [Fe/H] $\gtrsim$$-$1 are there any found with larger
amounts of Li.  There is a gradual increase in A(Li) with [Fe/H] in normal
dwarf stars to a maximum is near A(Li) of +3.3.  This results from the general
galactic enrichment in Li, Be, and B over time.  A large range in A(Li) of
three orders of magnitude can be seen in solar type stars caused by that
galactic enrichment and by the slow stellar Li destruction.

For both Be and B there is a marked and steady increase with [Fe/H].  Figures
14 and 15 show the clear increases in these elements over the course
of the evolution of the Galaxy.  The increase in A(Be) with [Fe/H] and [O/H]
(and with [Ti/H] and [Mg/Fe]) for 117 metal-poor stars emphasizes the trend
with chemical evolution (Boesgaard et al.~(2011).  

Some profound insights have come from the analysis of Li, Be, and B that are
related stellar interiors.  The surface abundances of these elements are
important guides to internal mixing.  The surface contents in clusters show
that light element depletion is a phenomenon of main sequence evolution.  The
dramatic drop in Li especially, but also Be and B, in the mid-F dwarfs is
apparently the result of extra internal mixing caused by rotation.  The
abundances of Li and Be are well correlated in main-sequence stars in clusters
and in the field over a span of a factor of 400 in Li abundance.  This is true
for Be and B in the Hyades and in field stars as well for a range of a factor
of 100 in Be.

\section{ADDITIONAL COMMENTS}

The classic image of the lone astronomer with his (sic) eye at the telescope
is far from true in today's world of astronomy.  The tools of our trade are
extremely complicated and require the talents and efforts of a genuine
multitude of people.  Just think of the complex business of designing and
building a modern telescope which is a multi-ton machine (300 tons for the
Keck telescopes) which has to track to better than 0.01 arcsec and has to
operate with better than clockwork precision.  It involves 24-inch I-beams
down to 00 screws.  Furthermore, that telescope needs sophisticated
instruments to detect photons from astronomical objects.  Those instruments
work best when accompanied by sensitive modern detectors.  And what would an
astronomer do without all the software designers who make the telescope
operate?  And the  ones who run the instruments that produce data?  And
those who provide technical support, even in the middle of the night?  There
are also those who devise the various tools that help in the analysis of that
data?  Even in the days of my youth when astronomers developed their own
photographic plates, they relied on data reduction techniques and equipment
developed by others.  At that time we had graphic artists and
photographers to make our plots and diagrams ready for publication.

My point here is that in order for astronomers to collect, analyze, and
interpret data from a scientific program they want to pursue, they rely on
virtual legions of other people.  Furthermore we interact with other
astronomers regarding our findings.  I am extremely grateful to this
collection of engineers, designers, computer experts, telescope operators,
assistants, colleagues and students.  Sometimes even small interactions
produce big thoughts or germinate new approaches to ongoing issues.  I want
them to know how aware I am of all their extensive help.  My deep thanks go to
all the many colleagues and students I've worked with and who have so expanded
my horizons.  I dedicate this talk and paper to Hans Boesgaard, my husband, my
best friend, my favorite telescope engineer.

\clearpage

\end{document}